\documentclass[aps,prb,twocolumn,reprint,showpacs,superscriptaddress,amssymb,floatfix]{revtex4-1}

\usepackage{bm}
\usepackage{graphicx}
\usepackage{latexsym}
\usepackage[dvips]{color} 

\begin{document}

\title{Quasiparticle-scattering measurements of laminar and turbulent vortex flow in the spin-down of superfluid $^3$He-B}

\author{J.J. Hosio} \email{jaakko.hosio@aalto.fi}
\affiliation{O.V. Lounasmaa Laboratory, P.O. Box 15100, FI-00076 AALTO, Finland}

\author{V.B. Eltsov}
\affiliation{O.V. Lounasmaa Laboratory, P.O. Box 15100, FI-00076 AALTO, Finland}



\author{M. Krusius}
\affiliation{O.V. Lounasmaa Laboratory, P.O. Box 15100, FI-00076 AALTO, Finland}

\author{J.T.  M\"{a}kinen}
\affiliation{O.V. Lounasmaa Laboratory, P.O. Box 15100, FI-00076 AALTO, Finland}

\date{\today}

\begin{abstract}

The dynamics of quantized vortices is studied in superfluid $^3$He-B after
a rapid stop of rotation. We use Andreev reflection of thermal excitations
to monitor vortex motion with quartz tuning fork oscillators in two different experimental setups at temperatures below $0.2T_\mathrm{c}$. Deviations from ideal cylindrical symmetry in the flow environment cause the early decay to
become turbulent. This is identified from a rapid initial overshoot in the
vortex density above the value before the spin-down and its subsequent decay
with a $t^{-3/2}$ time dependence. The high polarization of the vortices along the rotation axis significantly suppresses the effective turbulent kinematic viscosity $\nu'$ below the values reported for more homogeneous turbulence
and leads to a laminar late-time response. The laminar dissipation down to $T < 0.15T_\mathrm{c}$ is determined from the precession frequency of the polarized vortex configuration. In the limit of vanishing normal component density, it
is found to approach a temperature-independent value, whose origin is currently under discussion.

\end{abstract}


\maketitle 

\section{Introduction}

The dynamics of quantized vortices, especially superfluid turbulence at ultralow temperatures, where the vortices can move with little dissipation,
is an active topic of research.\cite{vinen} In the $T\rightarrow0$ limit, the superfluid can be modeled as an inviscid and incompressible fluid, where turbulence involves vortex reconnections and tangle formation. In these conditions, numerical modeling and the interpretation of measurements on turbulence become more manageable than in the case of classical viscous fluids.

One way to generate complex fluid motion is to bring a rotating container to rest. In classical fluids, the response of the fluid in such a spin-down experiment is governed by the frictional forces at the boundaries and becomes unstable at relatively low Reynolds numbers. \cite{mathis} Also in superfluid $^4$He, where vortex pinning at the container walls plays a significant role due to the small vortex core radius, an impulsive spin-down to rest is generally expected to generate turbulence. \cite{Manchester}

In superfluid $^3$He the situation is different, since in the case of smooth walls, the coupling to the container is accomplished mainly by the frictional interaction between the normal and superfluid component known as mutual friction. That is, the superfluid does not couple to the boundaries directly,
but instead, does so through volume forces to the normal fluid, which is stationary in the reference frame of the container. At high temperatures and large mutual friction vortex motion is laminar, but as the mutual-friction coupling vanishes exponentially in the zero-temperature limit, the flow of vortices is easily destabilized and becomes turbulent. Earlier studies in $^3$He-B show that in an axially symmetric environment the spin-down flow is laminar even in the limit of vanishing normal fluid density. \cite{PRL10} However, with increasing surface friction, the response ultimately becomes turbulent. \cite{AB_turb} Here we show that a similar change from a laminar to a turbulent spin-down response can be brought about by deviations in the flow geometry from axial symmetry. The ensuing  turbulence is concentrated in the initial phase of the response and since the polarization of the vortices along the symmetry axis remains high, laminar flow  ultimately takes over the late part of the response at lower vortex densities even at temperatures below $0.15T_{\rm c}$.

In superfluid  $^3$He-B, vibrating objects can be used to monitor the density of thermal excitations.
This has proven to be useful for studying different vortex structures, especially at ultralow temperatures, where practically all other detection methods become insensitive. \cite{fisher_pltp} The measurements rely on the fact that the rotating flow associated with quantized vortices can constrain the trajectories of the elementary excitations, which otherwise move along ballistic flight paths. In the presence of superfluid flow with velocity $\mathbf{v}_{\rm s}$ excitations with momentum $\mathbf{p}$ undergo the Galilean transformation $E\rightarrow E + \mathbf{p}\cdot\mathbf{v}_{\rm s}$. An excitation moving in the superfluid with insufficient energy to overcome the potential energy barrier created by the flow of vortices has no forward-propagating states and therefore retraces its trajectory changing flavor from quasiparticle to quasihole in a process called Andreev reflection. \cite{lanc93} Thus, the presence of vortices can be inferred from the variations in thermal damping of vibrating objects, such as vibrating wires or quartz tuning forks. The cross section of Andreev reflection is large even for sparse vortex structures providing a sensitive probe of the superfluid flow field. \cite{bradley, barenghi} Here we use these techniques to study the interplay of laminar and turbulent flow of vortices in the spin-down response when a deviation is introduced from ideal axially symmetric flow.

\section{Turbulent and laminar dynamics of superflow}

\subsection{Turbulent superflow}
Turbulence is often defined as a complex and dynamic flow field, which involves  processes spanning several orders of magnitude in spatial extent with aperiodic temporal dependence. In quantum turbulence, different types of turbulent flows are characterized by the local density $L(\mathbf{r})$ of the vortices and their polarization. In its simplest form, turbulence in superfluids consists of a homogeneous and isotropic tangle of singly quantized vortex lines.

On length scales large compared with the intervortex distance $\ell=L^{-1/2}$ quantum turbulence often resembles its classical counterpart. \cite{vinen} The energy is injected into eddies at length scales determined by the characteristic size of the flow disturbance. The large scale motion is achieved by partial polarization of vortices to bundles forming eddies of different sizes. As in classical turbulence, the energy cascades down with a Kolmogorov-type energy spectrum given by
\begin{equation}
E(k)=C\epsilon^{2/3} k^{5/3},
\label{kspec}
\end{equation}
where  $C\approx1.5$ is the Kolmogorov constant.
Assuming the dissipation is determined by the length scale $\ell$, the energy flux towards shorter length scales, i.e., inverse $k$, per unit mass is given by
\begin{equation}
\epsilon = \nu ' \kappa^2 L^2,
\label{eps}
\end{equation}
where $\nu '$ is the effective kinematic viscosity and $\kappa$ the circulation quantum. Using Eqs. (\ref{kspec}) and (\ref{eps}) allows us to describe the late-time decay of vortex density \cite{stalp} by
\begin{equation}
L = \frac{\sqrt{27C^3}D}{2 \pi\sqrt{\nu '} \kappa} t^{-3/2},
\label{dens}
\end{equation}
where the container size $D$ determines the cutoff wavenumber $k_0=2\pi/D$.
The value of $\nu'$ depends both on temperature and the nature of the flow.
Despite the fact that the derivation leading to Eq. (\ref{dens}) assumes homogeneous and isotropic turbulence, our experiments show that turbulence
with high polarization of vortices can also decay as $t^{-3/2}$. However, the results suggest that polarization of the turbulent vortex structure suppresses $\nu '$ significantly compared to the value measured for more isotropic turbulence.

It is widely believed that as the energy flows to length scales smaller than the intervortex distance, it is transferred in a cascade of helical deformations of individual vortices called Kelvin waves. \cite{vinen}
There is an ongoing debate on the nature of the energy transfer from the classical Kolmogorov-like cascade to the Kelvin-wave cascade. \cite{kozik,lvov,front1} At finite temperatures mutual friction provides dissipation at all length scales, and even at $T=0$, the energy is ultimately dissipated when Kelvin waves at very large $k$ induce phonon emission \cite{vinenprb} in $^4$He or perhaps quasiparticle emission \cite{silaev} from vortex cores in $^3$He-B.

\subsection{Laminar superflow}
The other extreme of dynamic vortex motion is fully laminar polarized flow supporting no vortex reconnections. In superfluid $^3$He-B the normal component is clamped to corotation with the cylindrical container with velocity ${\bf v}_{\rm n}=\Omega \hat{\bf z}\times \bf{r}$ in the cylindrical coordinates $(r,\phi,z)$  with $\Omega$ being the angular rotation velocity. At ultralow temperatures, where the reactive mutual friction can be neglected, the course-grained hydrodynamical equation for the superfluid velocity ${\bf v}_{\rm s}$ is given by \cite{sonin}
\begin{equation}
\frac{\partial {\bf v}_{\rm s} }{ \partial t}+ \nabla\mu- ({\bf v}_{\rm s} \cdot \nabla) {\bf v}_{\rm s}
 = - \alpha~\hat{\bf \omega} \times(( {\bf v}_{\rm s}- {\bf v}_{\rm n})  \times
(\nabla\times {\bf v}_{\rm s}) ),
\label{Hydrodynamics}
\end{equation}
where $\mu$ is the chemical potential and $\hat{\bf \omega}$ a unit vector along the vorticity $\nabla\times {\bf v}_{\rm s}$.
In the ballistic regime, the dissipative mutual-friction parameter $\alpha$ is proportional to the Boltzmann factor $\exp(-\Delta/T)$,
where $\Delta$ is the pressure-dependent superfluid energy gap.

If the vortices remain highly polarized along the axis of the cylinder, and thus vortex reconnections play no role, laminar rotating flow of the superfluid is solid-body-like with velocity ${\bf v}_{\rm s}=\Omega_s\hat{\bf z}\times \bf{r}$. In this case, taking a curl of both sides simplifies Eq. (\ref{Hydrodynamics}) to
\begin{equation}
\frac{d \Omega_s(t)}{dt}=2 \alpha \Omega(t)[\Omega(t)-\Omega_s(t)].
\label{de}
\end{equation}
For a step change of the rotation drive at $t=0$ from an angular velocity $\Omega_0$ to rest the solution of Eq. (\ref{de}) is given by
\begin{equation}
\Omega_s(t)=\frac{\Omega_0}{1+t/\tau},
\label{step}
\end{equation}
where $\tau=(2\alpha \Omega_0)^{-1}$. In real experiments the change of rotation velocity is done at a finite rate, in our case typically at $a=d\Omega/dt=-0.03$~rad/s$^2$. During the deceleration, i.e., for $-\Omega_0/a<t<0$, the solution of Eq. (\ref{de}) is
\begin{equation}
\Omega_s(t)=\frac{\sqrt{a}e^{\alpha (t+\Omega_0/a)(\Omega_0-at)}}{\tau_0\sqrt{a}+\sqrt{\pi \alpha}e^{{\alpha\Omega_0^2/a}}{\rm{erf}}(\sqrt{\alpha a}t)},
\label{finite}
\end{equation}
where $\tau_0=\Omega_0^{-1}+\sqrt{\alpha/a}\exp(\alpha\Omega_0^2/a){\rm{erf}}(\sqrt{\alpha/a}\Omega_0)$.
At low temperatures, where $\alpha\ll a/ \Omega_0^2$ the superfluid velocity at the end of the deceleration is very close to $\Omega_0$ and consequently, Eq. (\ref{step}) can be used at all times $t>0$.

\section{Measurement techniques}

\begin{figure}
\begin{center}
\centerline{\includegraphics[width=0.95\linewidth]{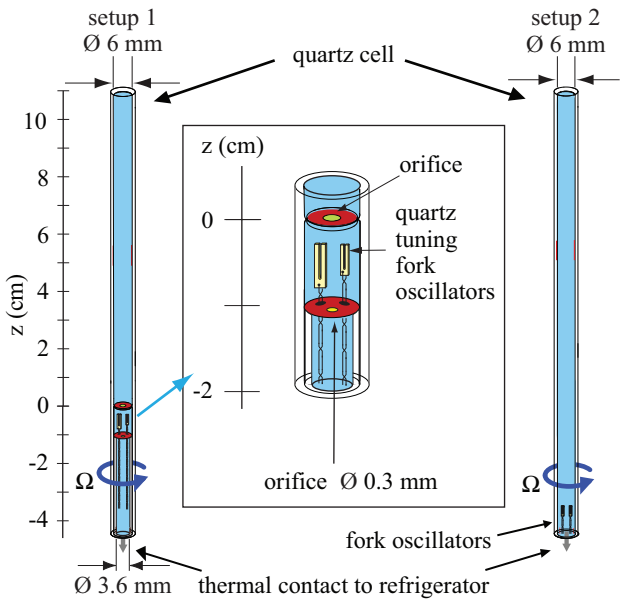}}
\caption{(Color online) (Left) Experimental setup for the bolometric measurement of the early turbulent spin-down response. The upper experimental volume modeled as a bolometer is separated from the heat exchanger volume at the bottom by the lower division plate with a small 0.3-mm-diameter orifice. The upper division plate with a 0.75-mm-diameter aperture secures laminar spin-down flow in the topmost section.The quartz tuning forks are calibrated to measure the heat transport to the refrigerator.
(Right) Setup without division plates for the measurement of the late laminar part of the spin-down response. }
\label{setup}
\end{center}
\vspace{-5mm}
\end{figure}

In our present measurements the $^3$He-B sample is contained in a smooth-walled fused quartz cylinder. Two experimental setups shown in Fig. \ref{setup} have been used. In setup 1 on the left the $^3$He-B sample is pressurized to 29~bar and the cylinder is divided in three parts by two 0.7-mm-thick quartz division plates with 0.75~mm and 0.3~mm orifices. The lowermost part consists of a 30-mm-long, 3.6-mm-inner-diameter tube, which opens at the bottom to the heat exchanger. This volume is cooled to the lowest possible temperature $T<0.14T_c$ with a nuclear demagnetization cooling stage. The two upper parts combined constitute a 12-cm-long smooth-walled section of the quartz tube with 6~mm inner diameter and act here as a bolometer. The top section above the orifice of 0.75 mm diameter has been used in earlier measurements for examining the stability of laminar vortex flow with NMR techniques. \cite{PRL10} The middle section located between the two division plates was added later to measure Andreev reflection from a cluster of rectilinear vortices \cite{hosio} and the thermal signal from vortex dissipation in turbulent spin-up. \cite{front} Here we use the tuning-fork oscillators in the middle section for measuring the spin-down response in the bottommost section. In this case vortex flow is perturbed at the bottom end of the sample tube  by the presence of a rough and wavy sintered heat-exchanger surface and the leads of the tuning forks.

In setup 2, shown on the right in Fig. \ref{setup} both division plates are removed and the sample is pressurized to 0.5~bar instead of 29~bar. In this setup, axial flow symmetry is disturbed by the presence of the two tuning forks placed at the bottom end of the quartz tube.

In both setups the container is initially filled with a uniform array of rectilinear vortices oriented along the rotation axis. To create the vortex array the cryostat is rotated at constant angular velocity $\Omega$ around the axis of the container tube. \cite{hosio} The vortex density in the equilibrium state is determined by minimization of the free energy in the rotating frame and is given by the solid-body-rotation value $L=2\Omega /\kappa$.

In the absence of applied heating, the temperature $T \approx 0.20T_c$ in the bolometer of setup 1 is determined by the background heat leak from the container walls ($\dot{Q}_{\rm{hl}}\approx10-20$~pW depending on rotation velocity) and the thermal resistance $R_{\rm T} (T)=(d\dot{Q}/dT)^{-1}$ from the tiny 0.3-mm orifice in the lower division plate. The thermal balance in the system is discussed in more detail in Ref. [\onlinecite{hosio}]. Assuming thermal equilibrium in the volume above the 0.3-mm orifice, all of the heat flux $\dot{Q}$ must leave through the orifice as a flux of energy-carrying excitations with the temperature dependence given by
\begin{equation}
\label{p2}
\dot{Q}(T)=\frac{4\pi k_B p_F^2}{h^3}Te^{-\frac{\Delta}{T}}(\Delta+k_B T) A_h.
\end{equation}
Here $A_h$ is the so-called effective area of the orifice, which depends on the number and orientation of the vortices below the 0.3~mm orifice. The reason for this is that due to the vortices, part of the thermal excitations experience Andreev reflection and thus, the effective area is reduced (see the inset of Fig. \ref{TL}). Therefore, information on vortex dynamics can be obtained from the temporal dependence of $A_h$.

The principle of the measurement is to study how the Andreev scattering from the vortices affects the fraction of thermal excitations returning to the bolometer,
\begin{equation}
f_{\rm r}(t)=1-A_h(t)/A_{h,0},
\label{refco}
\end{equation}
where $A_{h,0}$ is the effective area in the absence of vortices. The temperature is obtained from the resonance linewidth  $\Delta f$ of a quartz tuning fork oscillator, which  is proportional to the Boltzmann factor \cite{todo} $\exp(-\Delta/T)$ after subtracting the intrinsic linewidth $\Delta f_{\rm{int}}\approx15$~mHz. Thus, for known $\dot{Q}(T)$, $A_h(t)$ can be calculated directly from Eq. (\ref{p2}). The background heat leak $\dot{Q}_{\rm{hl}}$ in the steady state can be determined by measuring the temperature increase with one tuning fork as a function of a known heat input from the other fork. \cite{hosio}

\section{Bolometric measurement of turbulence in the spin-down response}
\label{turb_chap}
\begin{figure}
\begin{center}
\centerline{\includegraphics[width=0.98\linewidth]{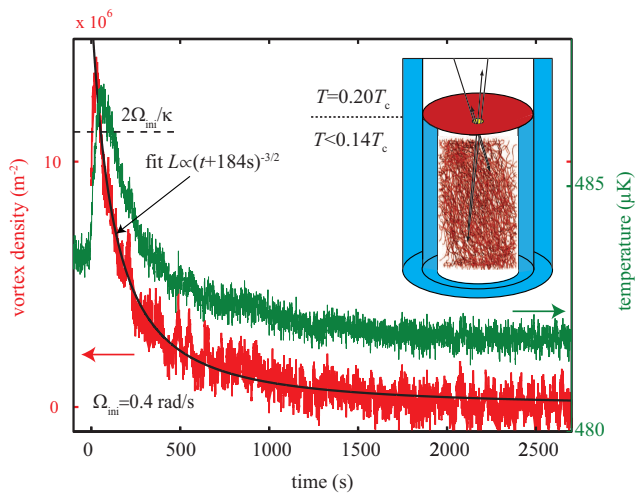}}
\caption{
(Color online) Temperature $T$ in the bolometer and an estimate for the lower limit of the vortex density $L$ after bringing the container to rest from $\Omega_{\rm{ini}}$=0.5~rad/s at $t=0$. The vortex density is inferred from the fraction of Andreev-reflected thermal excitations as discussed in the text. The solid line is a fit to $L\propto(t+\tau)^{-3/2}$ dependence after the initial overshoot in the data. The inset shows the principle of the measurement: The vortices below the orifice reflect some fraction of the thermal excitations back to the bolometer. The fraction depends on the density and configuration of these vortices at the lower temperature $T<0.14T_{\rm c}$. The measurement is performed at 29~bar liquid $^3$He pressure.}
\label{TL}
\end{center}
\vspace{-5mm}
\end{figure}

Figure \ref{TL} shows the temperature response of the bolometer to a rapid change of rotation. The moment $t=0$ corresponds to the time when the final constant
value $\Omega=0$ is attained after a change of drive at $\dot {\Omega}=-0.03$~rad/s$^2$ from the initial velocity $\Omega_{\rm{ini}}=0.5$~rad/s. The initial temperature increase during
the deceleration is a consequence of three different phenomena: direct heating from the decay of vortices inside the bolometer volume, weakly increased rotation-induced heat leak at certain velocities of mechanical resonances of the cryostat in the range $0<\Omega<\Omega_{\rm{ini}}$, and most importantly, increased thermal resistance $R_{\rm T}$ due to the spin-down-induced turbulence below the bolometer volume.

After the container is brought to rest, the heat generated in the bolometer $\dot{Q}_{\rm{gen}}(T)$, which is at $T\approx0.20T_{\rm c}$, is the sum of the known background heat leak and the heat release from the laminar decay of vortices stabilized by the mutual friction $\alpha\approx0.002$ in the upper sections of the cylinder above the 0.3~mm orifice. \cite{specif} The kinetic energy of the rotating superfluid with density $\rho_{\rm s}$ is given by
\begin{equation}
E_{\rm{kin}}=\pi \rho_{\rm s}  R^4 h \Omega_s^2/4,
\label{ekin}
\end{equation}
where $R$ and $h$ are the radius and the height of the cylindrical bolometer volume, respectively. Combining Eqs. ({\ref{step}) and (\ref{ekin}) yields the rate at which the kinetic energy translates into heat,
\begin{equation}
\dot{E}_{\rm{kin}}=\frac{\pi \rho_{\rm s}  R^4 h \Omega_0^2}{2\tau}(1+t/\tau)^{-3}.
\label{ekindot}
\end{equation}
Now the total heat flow out of the bolometer volume is given by
\begin{equation}
\dot{Q}(t)=\dot{Q}_{\rm{gen}}(t)-c(T)V\dot{T},
\label{heatflow}
\end{equation}
where $V=\pi R^2 h$ is the volume of the bolometer and the heat capacity $c(T)$ is given approximately by \cite{bauerle}
\begin{equation}
c(T)=k_B^2 \sqrt{2\pi}N_F\left(\frac{\Delta}{T}\right)^{\frac{3}{2}} e^{-\frac{\Delta}{T}}\left(\Delta+\frac{21}{16}T\right).
\label{hcap}
\end{equation}
Here, $N_F$ is the density of states at the Fermi level. From the measured temperature and its time derivative $\dot{T}$, one can solve for the reflection coefficient $f_{\rm r}(t)$ using Eqs. (\ref{p2}), (\ref{refco}), (\ref{heatflow}), and (\ref{hcap}).

\begin{figure*}
\begin{center}
\centerline{\includegraphics[width=0.98\linewidth]{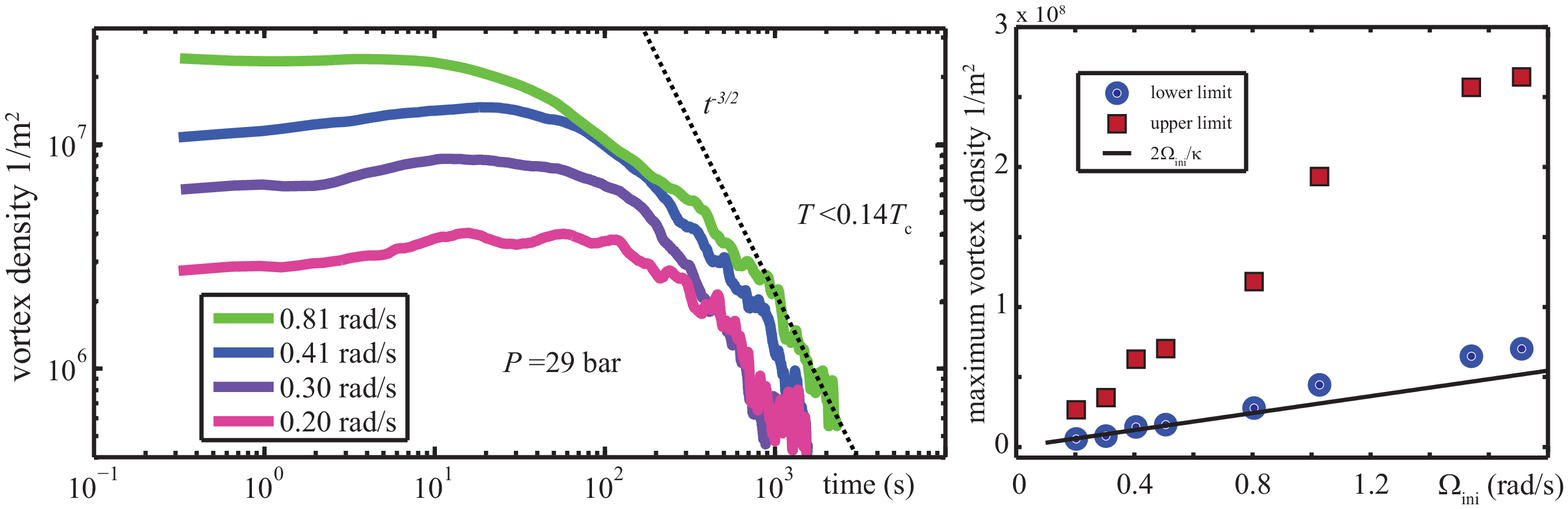}}
\caption{
(Color online) (Left) Lower limit of the vortex density $L$ as a function of time after bringing the container to rest for four initial velocities $\Omega_{\rm{ini}}$§ (see text for details). A simple moving average filtering is used to reduce the noise in the data. The upper limit of $L$ at each value of $\Omega_{\rm{ini}}$ obtained from a calibration measurement with rectilinear vortices at a known density is a few times higher. (Right) Lower and upper limits for the maximum vortex density generated in spin-down together with the initial steady-state value $2\Omega_{\rm{ini}}/\kappa$.}
\label{Llog}
\end{center}
\vspace{-5mm}
\end{figure*}
Even though we do not know the detailed structure of the vortex configuration below the 0.3~mm orifice reflecting a part of the thermal excitation beam back to the bolometer, two limiting cases can be considered to obtain upper and lower limits for the vortex density $L$. Obviously, the fraction of the Andreev-reflected beam at fixed $L$ is minimized if all of the vortices are polarized along the axis of rotation and are thus perpendicular to the plane of the bolometer orifice. Accordingly, the upper limit for the vortex density at a given reflection coefficient can be obtained from the measurements of $f_{\rm r}(\Omega)$ with known vortex density in uniform rotation (as measured in Ref. [\onlinecite{hosio}]). The lower limit can be estimated by considering a system of vortices which are perpendicular to the beam of excitations and have an intervortex spacing $\ell=L^{-1/2}$. In this case the vortex density is given by \cite{barenghi}
\begin{equation}
L=-\frac{2k_B T\ln(1-f_{\rm r})}{3\pi \Delta \xi S},
\label{Lper}
\end{equation}
where $\xi=2 \cdot 10^{-8}$~m is the coherence length and $S$ the spatial extent of vorticity. This simple estimate does not account for the diffusively scattering boundaries of the container. Our numerical simulations \cite{hosio} show that in the presence of vortices the wall effects increase $f_{\rm r}(\Omega)$ by almost exactly a factor of two. Thus, a reasonable lower limit
for $L$ is attained by dividing the measured reflection coefficient by two and taking $S$ to be the mean distance for excitations to travel before they would hit the boundary of the container in the absence of vortices. Figure \ref{TL} shows an example of the lower limit of $L$ calculated this way from the measured temperature response shown in the same figure. The upper limit obtained from the calibration measurements of straight vortices with known density is a few times larger.

The result for $L(t)$ in Fig. \ref{TL} carries the following two characteristic signatures from turbulence. First, an initial overshoot, of the order of a few tens of percents with the maximum at around $t=30$~s, indicates that part of the kinetic energy of the superfluid is converted to a turbulent tangle of vortices. The upper and lower limits of the maximum vortex density as a function of initial rotation velocity are shown in the right panel of Fig. \ref{Llog}. The lower limit maximum lies close to the known initial density $2 \Omega_{\rm{ini}}/\kappa$. Second, after a couple hundred seconds the estimated vortex density (both upper and lower limit) fits well with a $t^{-3/2}$ dependence on time. This suggests that at least some fraction of the vortices decays in a turbulent manner, since for a fully laminar response the decay should be proportional to $t^{-1}$, as given by Eq. (\ref{step}). In the left panel of Fig. \ref{Llog}, examples of the lower limit of the vortex density $L$ at different initial rotation velocities are plotted with logarithmic scales
illustrating the $t^{-3/2}$ time dependence. After $\sim 10^3$~s the vortex densities become too low to be detected with good accuracy.

The overshoot in Fig. \ref{TL} is roughly an order of magnitude smaller than in the more turbulent spin-down response of the measurements in superfluid $^4$He in a cubic container with grid-covered surfaces in Ref. [\onlinecite{Manchester}], but of similar magnitude as in the numerically calculated examples of more polarized turbulence in Refs. [\onlinecite{PRL10}] and [\onlinecite{AB_turb}]. The characteristic time to reach the maximum of the overshoot in  $L(t)$ decreases with increasing initial rotation velocity $\Omega_{\rm ini}$. With the present deceleration rate, the time to bring the cryostat to rest exceeds the time to reach the maximum level of $L(t)$ at high velocities.\cite{dec_com} In this situation, the line density remains approximately constant during the remaining deceleration.


It is tempting to relate the response in Fig. \ref{Llog} to Eq. (\ref{dens}) for the quasiclassical decay of the line density with the same $t^{-3/2}$ time dependence. One needs to be careful though, since some of the assumptions leading to Eq. (\ref{dens}), especially the homogeneity of the flow, are not met. Nonetheless, if the size of the energy containing length scale is chosen to be the diameter of the container, i.e. $D=2R$, then the effective kinematic viscosity $\nu '$ extracted from the lower limit of vortex densities is
$(1.0\pm0.3) \cdot 10^{-4} \kappa$ and does not depend on the initial rotation velocity in the range  $0.2\leq\Omega_{\rm{ini}}\leq1.7$~rad/s.
The estimate obtained from the upper limit is roughly an order of magnitude smaller.

Earlier experiments on the free decay of turbulence in  $^3$He-B in the $T=0$ limit have been done in a series of measurements by Bradley et al. They used a vibrating grid to create a tangle of vortices and inferred the vortex density from a reduction in the damping of vibrating wire resonators caused by Andreev
reflection from the flow field of the vortices. \cite{lanc06} In their experiment, the decay was more than an order of magnitude faster and yielded $\nu ' \approx 0.3 \kappa$, which is two orders of magnitude larger than the value measured in $^4$He for the free decay of tangles produced by rapid spin-down. \cite{Manchester} The orders-of-magnitude-smaller value for the effective kinematic viscosity in our experiment underlines the influence of the large polarization of vortices, which suppresses the amount of vortex reconnections and thus dissipation in comparison to more homogeneous turbulence.

Since the reflection coefficient is more sensitive to the vortices which are perpendicular to the excitation beam than to those parallel to it, more quantitative estimates of $L$ cannot be done without detailed information on the polarization of the vortex structure. Supposedly, in the beginning the vortex configuration is more randomized and less polarized along the rotation axis. These vortices reconnect with larger probability than the polarized vortices and account for the observed $t^{-3/2}$ dependence in the early part of the decay. This assumption is supported both by numerical \cite{JLTP10}
and experimental (Sec. \ref{osc_chap}) results.

\section{Laminar Decay inferred from a precessing vortex cluster}
\label{osc_chap}
In the absence of the division plates in setup 2, the quartz tuning-fork resonators can be used to probe the spin-down-induced flow field locally.
In the equilibrium vortex state at constant rotation, the ballistic nature of thermal excitations assures thermal equilibrium inside the container. Dynamic inhomogeneous vortex structures, however, can shadow a part of the background excitation flow emanating from the walls, and thus, produce local temperature variations. Recently, these variations, measured with vibrating wires, were used to study line-density fluctuations in a turbulent vortex tangle. \cite{lanc08}

Here we study the response of the superfluid, originally rotating at constant angular velocity $\Omega_s=\Omega_{\rm{ini}}$, to a rapid spin-down to rest
by following the variation in the thermal excitation density in the vicinity of a tuning-fork oscillator. The background temperature in the range $150-190\,\mu$K is controlled by magnetization and demagnetization of the nuclear cooling stage.

\begin{figure}
\begin{center}
\centerline{\includegraphics[width=0.98\linewidth]{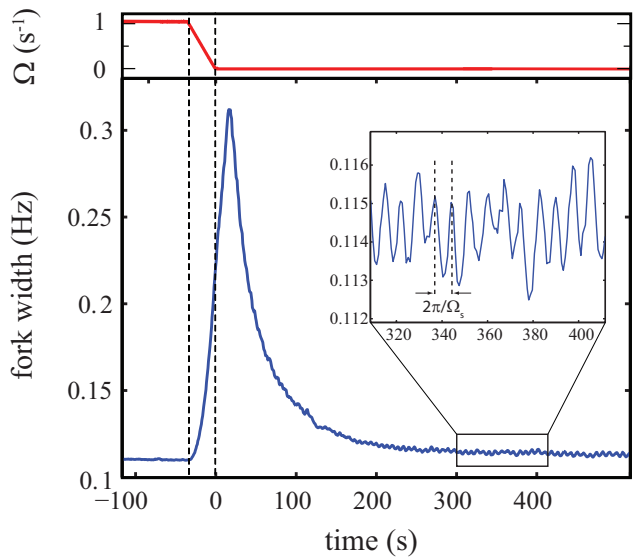}}
\caption{ (Color online) Response of the resonance width of the quartz tuning fork oscillator to a rapid spin-down from $\Omega_{\rm{ini}}$=1.02 rad/s to rest. The initial increase arises mainly owing to Andreev reflection from the rapidly decaying turbulent vortex configuration generated in the early part of the spin-down. The inset shows the zoomed view of the oscillations in the  laminar late response caused by a periodic variation of the thermal excitation density in the vicinity of the fork. These oscillations result from the Andreev scattering from an asymmetric vortex cluster precessing at angular velocity $\Omega_s(t)$. The measurement is performed at 0.5~bar liquid $^3$He pressure.}
\label{fig:osc}
\end{center}
\vspace{-5mm}
\end{figure}
Figure \ref{fig:osc} shows an example of the fork-oscillator response before, during, and after bringing the container to rest. The initial increase of the linewidth arises mainly from Andreev scattering from the turbulent vortex configuration generated in spin-down as described in Sec. \ref{turb_chap}. After about 200~s, periodic oscillations can be observed on top of the relaxing average temperature. We interpret these oscillations to originate from
a cluster of predominantly straight vortices precessing at the superfluid angular velocity $\Omega_s$. Other experimental signals of a similar type from precessing vortex bundles have been reported in Refs. [\onlinecite{front}] and [\onlinecite{risto}]. Here a small asymmetry in the structure of the cluster causes periodic variation of the thermal excitation density in the vicinity of the tuning fork oscillator. The frequency of these oscillations drops as $t^{-1}$ for several hours indicating that the cluster decays in a laminar manner. Using $\Omega_0$ and $\tau$ as fitting parameters produces an excellent fit to Eq. (\ref{step}) with $\Omega_0=b \Omega_{\rm{ini}}$ and $\tau \propto \Omega_0^{-1}$ as expected for solid-body-like laminar decay, illustrated in the right panel of Fig. \ref{fig:osc}. Measurements at different initial rotation velocities $\Omega_{\rm{ini}}=0.6-1.5$~rad/s and temperatures
$T=0.15-0.19~T_{\rm c}$ show that the constant $b\approx0.8$ and depends neither on $\Omega_{\rm{ini}}$ nor temperature.

\begin{figure}
\begin{center}
\centerline{\includegraphics[width=0.98\linewidth]{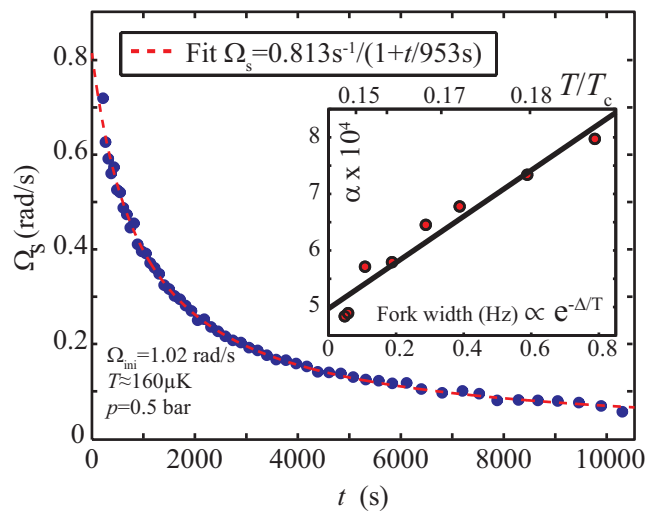}}
\caption{ (Color online) Superfluid angular velocity $\Omega_s$ as a function of time after bringing the container to rest at $t=0$. The response fits perfectly to the laminar flow model [Eq. (\ref{step})] shown by the dashed line with $\Omega_0\approx0.80 \Omega_{\rm{ini}}$ and $\alpha \approx 6.45 \cdot 10^{-4}$. The inset shows the dissipative mutual-friction parameter $\alpha$ as a function of the linewidth of the tuning-fork oscillator (bottom axis) or temperature (top axis). The value of $\alpha$, whose uncertainty is of the order of $\pm 5\cdot10^{-5}$, is extracted from the precession frequency of the vortex cluster, as discussed in the text, and averaged over $2-4$ measurements at different rotation velocities in the range $\Omega_{\rm{ini}}=0.6-1.5$~rad/s at the given temperature.}
\label{per}
\end{center}
\vspace{-5mm}
\end{figure}
Both fitting parameters provide valuable information on the vortex dynamics during the spin-down. The rapid drop of the superfluid angular velocity from $\Omega_{\rm{ini}}$ to $\sim 0.8 \Omega_{\rm{ini}}$ indicates that during the turbulent early part of the decay about one fifth of the vortices are annihilated due to vortex reconnections. The initial turbulent part seems to last a shorter time than in the measurements at 29~bar in Fig. \ref{Llog}.
Two reasons for the increased turbulence in the high-pressure measurements can be given: At 0.5~bar pressure the vortex core radius is roughly five times larger than at 29~bar. Therefore, vortex pinning on the surface of the heat exchanger, which is sintered from $\sim 10$-$\mu$m-size copper flakes, is expected to be stronger at high pressures. Second, the short narrow-diameter bottom part of the cylinder in setup 1 is markedly different from that of setup 2.

The decay of the remaining polarized vortices is dominated by the mutual friction with the parameter $\alpha=(2 \tau \Omega_0)^{-1}$. To date, the parameter $\alpha$ at low pressures has been measured only at temperatures above 0.35~$T_{\rm c}$, \cite{bevan} where the normal fluid density is three orders of magnitude higher than at the lowest temperature of this work. Eltsov et al. \cite{PRL10} used NMR-techniques to measure $\alpha$ down to 0.20~$T_{\rm c}$ at high pressures.

The vortex mutual friction is a consequence of the interaction between the thermal and vortex-core-bound excitations. \cite{kopnin} Since the normal-fluid density in the ballistic regime drops exponentially, the mutual-friction parameter should be proportional to the linewidth of the tuning-fork oscillator, $\Delta f \propto \exp(-\Delta/T)$. The inset of Fig. \ref{per} demonstrates
how $\alpha$ follows the expected exponential dependence on temperature, but
with a non-zero intercept $\alpha(0)\sim5\cdot10^{-4}$ in the $T\rightarrow0$ limit. The source of this zero-temperature dissipation is unclear. One possibility is surface interactions with the container walls. Another recently suggested mechanism is local heating of the vortex cores in accelerating motion, \cite{silaev} which can lead to temperature-independent dissipation in the zero-temperature limit. Finite zero-temperature dissipation has been observed earlier in the turbulent front propagation. \cite{front1} Whether the origin of the finite value for $\alpha$ is the same here as in the front motion is uncertain and remains to be clarified.

\section{Conclusions}
In conclusion, we have studied the spin-down response of superfluid $^3$He-B in a cylindrical container at very low temperatures, where the normal fluid excitations are extremely dilute. The low mutual friction together with small deviations from an ideal axially symmetric flow environment results in turbulence. This is in contrast to our earlier measurements at higher temperatures $T\geq0.2T_{\rm c}$ with more symmetric geometry. \cite{PRL10}

Vortex reconnections concentrate in the early part of the decay where they cause loss of polarization, increased dissipation, and $t^{-3/2}$ time dependence of the vortex density. Similarly to increased surface friction in Ref. [\onlinecite{AB_turb}], the increased vortex pinning on the rough and wavy heat-exchanger surface leads to increased turbulence also. The vortex core size seems to affect the decay, which agrees with the conclusion that vortex pinning plays a role in maintaining turbulent flow.

The vortex density in the late decay can be inferred with high accuracy from the local temperature variations caused by Andreev reflection from the precessing slightly asymmetric cluster of quantized vortex lines. This novel method reveals that the late decay of the vortex cluster is laminar even at the lowest temperatures. The temporal dependence of the spin-down decay allows us to determine both the amount of vortices lost in the initial turbulent part of the decay and the vortex mutual friction down to much lower normal fluid densities than before.

The dissipative mutual friction parameter $\alpha$ is found to approach a small nonzero value in the low temperature limit. This is the first time this type of residual friction is observed for laminar motion. Whether it signals true zero-temperature dissipation or some nonexponentially decaying dissipative
phenomenon remains as an interesting challenge for the future.

\section*{ACKNOWLEDGMENTS}
This work is supported by the Academy of Finland (Centers of Excellence Programme 2006-2011 and 2012-2017) and the EU 7th Framework Programme (FP7/2007-2013, grant 228464 Microkelvin). JH acknowledges financial support from the V\"{a}is\"{a}l\"{a} Foundation of the Finnish Academy of Science and Letters.


\begin{thebibliography}{99}
\bibitem{specif} In the small lower part of the bolometer, between the division plates, the response is not fully laminar due to the presence of the tuning forks, which obstruct the flow. This volume, however,
stores only $\sim10\%$ of the initial kinetic energy in the bolometer. We approximate the response to be laminar in the whole bolometer volume.
\bibitem{dec_com} At the lowest temperatures in $^3$He-B, the requirements for the rotation-induced heat leaks are more demanding than in $^4$He.
This limits the maximum decelaration rate to $\dot{\Omega}\sim-0.03$~rad/s$^2$.
\end{thebibliography}
\end{document}